\documentclass[a4paper,fleqn,usenatbib]{mnras}

\usepackage[T1]{fontenc}
\usepackage{ae,aecompl}

\usepackage{graphicx}	
\usepackage{amsmath}	
\usepackage{amssymb}	

\title[Orbit of HD 26441]{Precise orbital elements, masses and parallax of
  the spectroscopic--interferometric binary HD 26441}

\author[J. A. Docobo et al.]{
Jos\'e A. Docobo,$^{1}$\thanks{E-mail: joseangel.docobo@usc.es (JAD)}
Roger F. Griffin,$^{2}$
Pedro P. Campo,$^{1}$
Ahmad A. Abushattal$^{1}$
\\
$^{1}$Observatorio Astron\'omico Ram\'on Mar\'ia Aller. Universidade de Santiago de Compostela. \\
\hskip1.5em Avda. das Ciencias sn, Campus Vida, 15782 Santiago de Compostela, Spain\\
$^{2}$The Observatories, Institute of Astronomy, Madingley Road, Cambridge CB3 0HA
}

\date{Accepted 2017 April 10. Received 2017 April 6; in original form 2017 March 8}

\pubyear{2017}

\begin{document}
\label{firstpage}
\pagerange{\pageref{firstpage}--\pageref{lastpage}}
\maketitle

\begin{abstract}
The orbit of the visual--speckle binary  A 2801 (HD 26441) was calculated independently of the previously published double-lined spectroscopic solution, obtaining similar values for the common orbital elements. In this way, it has been possible to provide a consistent 3D orbit and precise values for both the orbital parallax and the individual masses of the components, so the physical properties of this stellar system are determined.
\end{abstract}




\section{Introduction}

Double-lined spectroscopic binaries that are optically resolved with 
high-resolution techniques are a fundamental source of information in stellar
astronomy. Indeed, once both orbits have been determined with high
precision, the spectroscopic one by means of radial velocities and the
visual one from the relative positions, we can obtain not only the
three-dimensional orbit of the system but also the individual masses of the
components as well as the orbital parallax. Such parallaxes are important because they may be used to check the values obtained by astrometric space missions, such as Gaia.

Since speckle interferometry \citep{1970A&A.....6...85L} was applied to
double-star astronomy, it has created a revolution in the precision of
the measurements.  Moreover, several authors
\citep{1976PASP...88..957M,1977ApJ...212..459M,1978ApJ...223..526M,1984SvAL...10...95B,1984A&AS...57...31B,1986A&AS...65...27B}
have used it to resolve binaries the study of which had previously been
restricted to radial velocities. The combination of the two techniques yields
values for parameters that are essential for the investigation of stellar
evolution.

Along this line of work we present the HD 26441 \linebreak (A 2801;  ADS 3041;
WDS 04107\hspace{0.5pt}$-$0452) system, the orbit of which, as a double-lined spectroscopic
binary, was obtained by Griffin \citep{2015Obs...135..321G} with a
period of 7527 days.

Although in this case the binary had been resolved by means of micrometer
measurements for nearly a century, and in 1961 with an optical interferometer
by Finsen, it was only in 1996 that the first speckle observations were carried
out by E. Horch. The micrometric measurements are summarized in
Table~\ref{tab:micmeas}, in which the first three columns correspond to the
date, position angle (in degrees), and separation (in arcseconds). The fourth column shows the observed apparent magnitudes of both stars (C. E. Worley reported $\Delta$m = 0 in his observation of 1975.720), and the fifth, the number of observing nights. The sixth column includes the observer code,
and finally, the seventh, the telescope aperture (in metres). The observer codes correspond to: B, van den Bos; VOU, Voute; VBS, Van Biesbroeck; MUL, Muller; HLN, Holden; WOR, Worley.

Table~\ref{tab:intmeas} lists the interferometric measurements, taken from
the {\it Fourth Catalog of Interferometric Measurements of Binary Stars}
\citep{2001AJ....122.3480H}. The first five columns include the date, the
position angle and separation (with uncertainties when available). Column six shows the difference in magnitude and its uncertainty.  A colon after the measurement indicates a dubious value. E. P. Horch indicated that the uncertainty of $\Delta$m is 0.12 in his observation of 2008.7024. On the other hand, the combined magnitude (7.50) is only available from Hipparcos. The seventh and eighth columns correspond to the filter used and its full-width half maximum in nanometres. The rest of
the columns are the aperture of the telescope (in metres), the number of
observing nights, the original reference and the code of the technique
(following the format of the catalog). The reference codes are as follows:
FIN61, \citet{1961CiUO..120..367F}; ISO90, \citet{1990PNAOJ...1..217I};
PER97, \citet{1997AA...323L..49P}; HOR01, \citet{2001AJ....121.1597H};
HOR12, \citet{2012AJ....143...10H}; TOK10, \citet{2010AJ....139..743T};
TOK12, \citet{2012AJ....143...42H}; RID15, \citet{2015ApJ...799....4R};
TOK14, \citet{2014AJ....147..123T}; TOK15, \citet{2015AJ....150...50T};
HOR15, \citet{2015AJ....150..151H}; TOK16, \citet{2016AJ....151..153T}. The techniques correspond to: J, visual interferometry; S, speckle interferometry; Hh, \linebreak Hipparcos measurement; St, Tokovinin speckle camera (SOAR); Ag, laser-guided adaptive optics.

\begin{table}
	\centering
	\caption{Micrometric measurements.}
	\label{tab:micmeas}
	\begin{tabular}{lrccclc}
		\hline
		Date & \multicolumn{1}{c}{$\theta$} & $\rho$ & Mag. &  n & Obs. & Tel.\\
		 & \multicolumn{1}{c}{($^{\circ}$)} & ($''$) &  &  &  & (m.)\\		
		\hline
		1936.33 & \hspace{8.5pt}2.8 & 0.18 & 8.3--8.3 & 3 & B & 0.7\\
		1937.68 & 183.1 & 0.20 & 8.0--8.0 & 4 & VOU & 0.6\\
		1938.68 & 184.1 & 0.19 &  & 4 & VOU & 0.6\\
		1941.07 & \hspace{4pt}21.7 & 0.17 & 8.0--8.0 & 1 & B & 0.7\\
		1944.33 & 202.9 & 0.17 & 8.2--8.2 & 2 & VBS & 2.1\\
		1944.72 & 216.3 & 0.16 &  & 4 & VOU & 0.6\\
		1953.73 & 168.8 & 0.12 &  & 2 & MUL & 0.9\\
		1955.01 & 354.7 & 0.14 &  & 6 & VBS & 2.1\\
		1955.52 & 356.0 & 0.18 & 8.2--8.3 & 2 & B & 0.7\\
		1956.17 & 357.7 & 0.19 &  & 3 & MUL & 0.6\\
		1958.03 & \hspace{8.5pt}5.1 & 0.18 & 8.4--8.5 & 1 & B & 2.1\\
		1958.09 & 357.3 & 0.20 & 8.4--8.6 & 1 & B & 2.1\\
		1959.14 & \hspace{4pt}12.2 & 0.14 & 8.2--8.2 & 2 & VBS & 2.1\\
		1959.967 & \hspace{4pt}12.1 & 0.19 &  & 1 & VBS & 1.0\\
		1960.192 & \hspace{4pt}11.8 & 0.16 &  & 1 & VBS & 2.1\\
		1961.77 & \hspace{4pt}15.0 & 0.17 & 8.3--8.5 & 4 & B & 0.9\\
		1972.920 & 183.3 & 0.20 &  & 1 & HLN & 1.0\\
		1975.720 & 348.3 & 0.16 &  & 3 & WOR & 1.5\\
		\hline
	\end{tabular}
\end{table}

\begin{table*}
	\centering
	\caption{Interferometric measurements.}
	\label{tab:intmeas}
	\begin{tabular}{lrlllclccclc}
		\hline
		\multicolumn{1}{c}{Date} & \multicolumn{1}{c}{$\theta$} & \multicolumn{1}{c}{$\sigma_{\theta}$} & \multicolumn{1}{c}{$\rho$} & \multicolumn{1}{c}{$\sigma_{\rho}$} & \multicolumn{1}{c}{$\Delta$m} & \multicolumn{1}{c}{Filter} & \multicolumn{1}{c}{FWHM} & \multicolumn{1}{c}{Tel.} & \multicolumn{1}{c}{n} & \multicolumn{1}{c}{Ref.} & \multicolumn{1}{c}{Tech.}\\
		 & \multicolumn{1}{c}{($^{\circ}$)} & \multicolumn{1}{c}{($^{\circ}$)} & \multicolumn{1}{c}{($''$)} & \multicolumn{1}{c}{($''$)} &  & \multicolumn{1}{c}{(nm.)} & \multicolumn{1}{c}{(nm.)} & \multicolumn{1}{c}{(m.)} &  &  & \\
		\hline
		1961.09 & 21.3 &  & 0.180 &  & 0.0 &  &  & 0.7 & 1 & FIN61 & J\\
		1988.8030 & \multicolumn{4}{c}{unresolved} &  &  &  & 2.1 & 1 & ISO90 & S\\
		1988.8057 & \multicolumn{4}{c}{unresolved} &  &  &  & 2.1 & 1 & ISO90 & S\\
		1991.25 & \multicolumn{4}{c}{unresolved} &  & 511 & 222 & 0.3 & 1 & PER97 & Hh\\
		1996.8984 & 355.5 &  & 0.169 &  &  & 503 & \hspace{4pt}40 & 2.1 & 1 & HOR01 & S\\
		2008.7024 & 47.5 & 1.4 & 0.100 & 0.003 & 0.52 & 550 & \hspace{4pt}40 & 3.5 & 1 & HOR12 & S\\
		2008.7702 & 47.1 & 0.2 & 0.0995 & 0.0003 & 0.8 & 551 & \hspace{4pt}22 & 4.1 & 1 & TOK10 & S\\
		2010.8920 & 76.0 & 0.3 & 0.0685 & 0.0005 & 0.0: & 788 & 132 & 4.1 & 2 & TOK12 & St\\
		2010.8920 & 76.5 & 0.3 & 0.0548 & 0.0003 & 0.9 & 534 & \hspace{4pt}22 & 4.1 & 2 & TOK12 & St\\
		2010.9657 & 75.6 & 0.3 & 0.0707 & 0.0002 & 0.9 & 534 & \hspace{4pt}22 & 4.1 & 2 & TOK12 & St\\
		2012.7572 & \multicolumn{4}{c}{unresolved} &  & 754 & 119 & 1.5 & 1 & RID15 & Ag\\
		2013.7369 & 146.4 & 0.3 & 0.0386 & 0.0001 & 0.7 & 534 & \hspace{4pt}22 & 4.1 & 2 & TOK14 & St\\
		2014.0457 & \multicolumn{4}{c}{unresolved} &  & 788 & 132 & 4.2 & 1 & TOK15 & St\\
		2014.0457 & \multicolumn{4}{c}{unresolved} &  & 534 & \hspace{4pt}22 & 4.2 & 2 & TOK15 & St\\
		2014.7583 & 333.9 &  & 0.0731 &  & 0.66 & 692 & \hspace{4pt}40 & 4.3 & 1 & HOR15 & S\\
		2014.7583 & 333.9 &  & 0.0725 &  & 0.67 & 880 & \hspace{4pt}50 & 4.3 & 1 & HOR15 & S\\
		2014.8538 & 338.7 & 0.6 & 0.0788 & 0.0009 & 0.7 & 788 & 132 & 4.2 & 2 & TOK15 & St\\
		2015.0286 & 337.3 & 0.1 & 0.0875 & 0.0002 & 0.6 & 788 & 132 & 4.2 & 2 & TOK15 & St\\
		2015.7437 & 344.3 & 0.0 & 0.1199 & 0.0005 & 0.7 & 788 & 132 & 4.1 & 2 & TOK16 & St\\
		\hline
	\end{tabular}
\end{table*}

Several visual orbits have been calculated for this system. The first one
\citep{1954AJ.....59..388M}, with a 20-year period, was revised later by
\citet{1972IAUDS.56R...1D} and \citet{1986AAS...65..551B} who proposed circular
orbits with around 40-year periods. All three of them show very large
residuals for the interferometric measurements. \citet{2014AJ....147..123T}
recovered the short-period eccentric orbit (P = 20.42 yr, e = 0.887) but it
does not fit the latest measurements, which justifies a revision.

In this article, we present a three-dimensional solution for the orbit of
the system based on a new astrometric orbit with a very good fit to the
micrometric measurements and remarkably to the speckle ones and which, at the
same time, agrees with the spectroscopic orbit on its common elements.  In
this way, it has been possible to present values for the masses of the
components with small uncertainties and also a value for the orbital parallax
which is a perfect match to that obtained by the {\it Hipparcos\/} mission.  
At present, the possible {\it GAIA\/} parallax has not been published.

A study of the physical properties of HD 26441 system completes this work.

\section{The new astrometric orbit}

In 2014, a methodology was published \citep{2014AstBu..69..461D} that
allows a determination of the orbit of a spectroscopic binary when we know the
parallax and a precise astrometric measurement ({\it i.e.} speckle). 
That methodology also has an important application which is analyzing the
coherence of the speckle measurements (and all the astrometry, in general)
by means of its adaptation to the spectroscopic orbit. In that way, it is
possible to improve on a previous study to find out which speckle measurements
should have more weight.

Following that procedure, it turned out that the speckle observations
performed in 2008.7024 (Horch), 2013.7369 (Tokovinin) and 2015.0286
(Tokovinin) are the ones which fit best. They are chosen in order to
use J.A. Docobo's analytic method
\citep{1985CeMec..36..143D,2012ocpd.conf..119D} and to generate a set of
relative orbits whose corresponding apparent orbits pass through the three
selected points.

Once the set of orbits is obtained, it suffices to choose that with a better rms
in $\theta$ and $\rho$ (taking into account also the arithmetic means, AM, of the
said coordinates) with the rest of the observations. The selected orbit,
which has the best fit with those requisites is very consistent, in the elements that they have in
common, with the spectroscopic orbit obtained by Griffin. The weighting scheme for the micrometer measurements is described
in \citet{2003AA...409..989D} and, for the interferometric ones, we assign weight
5 for telescopes with an aperture of less than 1m, 10 for 1--2m-class telescopes,
15 for 2--4m, and 20 for 4m and over.  The micrometer measurement of 1972.920 was not
taken into account in the calculations as it shows high residuals for all
the visual orbits.

\begin{figure}
	\includegraphics[width=\columnwidth]{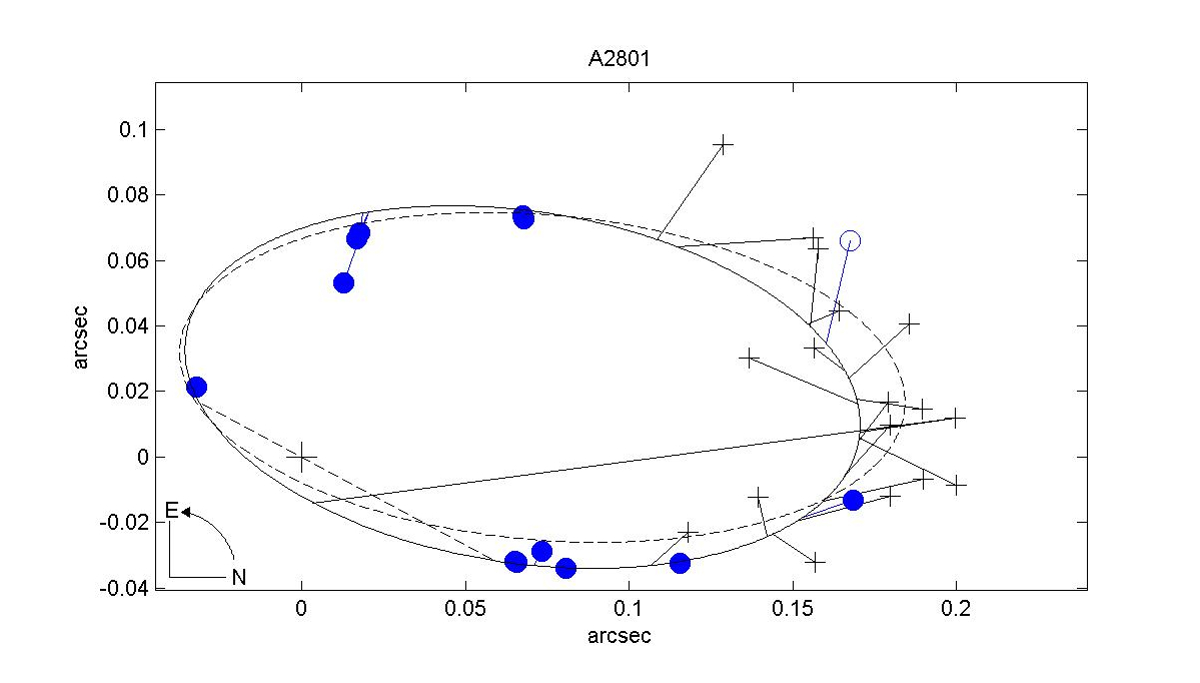}
    \caption{New (solid line) and previous (dashed line) orbit of HD 26441. Plusses represent visual observations, open circles correspond to eyepiece interferometry, and filled circles stand for speckle interferometry.}
    \label{fig:visorb}
\end{figure}

The new orbital elements with their uncertainties and quality indicators
(RMS and AM for \hbox{$\theta$ and $\rho$)} are provided in
Table~\ref{tab:astorb}. The numbers between brackets in the RMS and AM
correspond to the previous orbit by Tokovinin et al. The plot of the new
(solid line) and previous (dashed line) orbits can be seen in
Figure~\ref{fig:visorb}. The observations follow the 6th Orbit Catalog
\citep{2001AJ....122.3472H} convention: plusses, open circles, and filled
circles represent visual observations, eyepiece interferometry, and speckle
interferometry, respectively. 

Two new measurements that have recently been published \citep{2017arXiv170306253H} are also in good
agreement with our orbit.

\begin{table}
	\centering
	\caption{New astrometric orbit.}
	\label{tab:astorb}
	\begin{tabular}{ll}
		\hline
		Element & Value $\pm$ $\sigma$ \\
		\hline
		P (yr) & 20.616 $\pm$ 0.015\\
		T & 2014.035 $\pm$ 0.002\\
		e & 0.843 $\pm$ 0.002\\
		a ($''$) & 0.161 $\pm$ 0.002\\
		i ($^{\circ}$) & 66.4 $\pm$ 0.5\\
		$\Omega$ ($^{\circ}$) & 331.8 $\pm$ 0.5\\
		$\omega$ ($^{\circ}$) & 248.7 $\pm$ 0.5\\
		 & \\
		RMS$_{\theta}$ ($^{\circ}$) & 2.315 (6.853)\\
		RMS$_{\rho}$ ($''$) & 0.011 (0.017)\\
		 & \\
		AM$_{\theta}$ ($^{\circ}$) & 0.938 ($-$4.188)\\
		AM$_{\rho}$ ($''$) & $-$0.001 ($-$0.011)\\
		\hline
	\end{tabular}
\end{table}

\section{The spectroscopic orbit}

HD 26441 was placed on the Cambridge radial-velocity observing programme by RFG in 1986; 17 observations were made on a guest-investigator basis at
Haute-Provence and two at ESO, and 27 have been made in recent years with
the Cambridge `Coravel' spectrometer. Two periastron passages, when the
`dips' given by the two components were clearly separated, have been
observed. A trace obtained at Cambridge, that well illustrates the
unequal double-lined `dip', is shown here as Fig.~2. A more comprehensive
description of the radial-velocity observing campaign, and the derivation
of the spectroscopic orbit with the help of 37 additional observations
made by other observers with the Haute-Provence `Coravel' and kindly
subscribed for our use by the observers concerned, has already been
published in a paper \citep{2015Obs...135..321G} to which the interested reader's
attention is drawn. The orbital elements are included in Table~\ref{tab:sporb}, where they may be compared with the (entirely
independent) astrometric elements.

\begin{table}
	\centering
	\caption{The spectroscopic orbital elements.}
	\label{tab:sporb}
	\begin{tabular}{ll}
		\hline
		Element & Value $\pm$ $\sigma$ \\
		\hline
		P (d) & 7527 $\pm$ 4 (20.608 $\pm$ 0.011 yr)\\
		T (MJD) & 56670.5 $\pm$ 0.9 (2014.037 $\pm$ 0.003 )\\
		$\gamma$  (km/s) & +26.55 $\pm$ 0.03\\
		K$_{1}$ (km/s) & 11.59 $\pm$ 0.06\\
    K$_{2}$ (km/s) & 12.51 $\pm$ 0.10\\
		q & 1.079 $\pm$ 0.010\\
		e & 0.8372 $\pm$ 0.0015\\
		$\omega_{1}$ ($^{\circ}$) & 69 $\pm$ 0.4\\
    a$_{1}\;$sin$\:$i (Gm) & 656 $\pm$ 4\\
    a$_{2}\;$sin$\:$i (Gm) & 708 $\pm$ 6\\
    a$\;$sin$\:$i (Gm) & 1364 $\pm$ 8\\
    $\mathcal{M}_{1}$sin$^{3}$i (M$_{\odot}$) & 0.929 $\pm$ 0.021\\
    $\mathcal{M}_{2}$sin$^{3}$i (M$_{\odot}$) & 0.861 $\pm$ 0.016\\
		\hline
	\end{tabular}
\end{table}

\begin{figure}
	\includegraphics[width=\columnwidth]{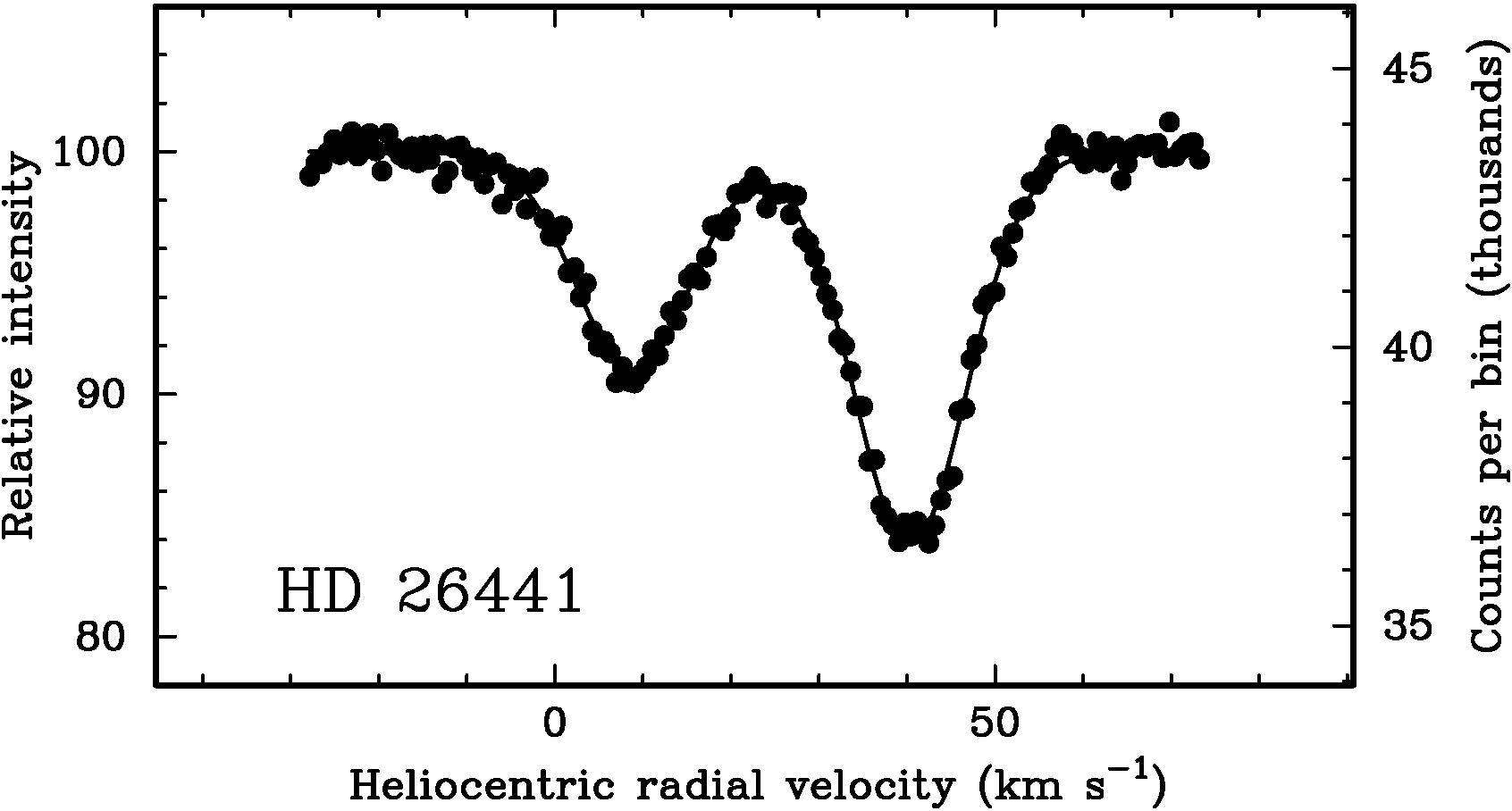}
    \caption{Radial-velocity trace of HD 26441, obtained with the Cambridge `Coravel' on
2013 October 17 and showing the two dips at practically their maximum separation.}
    \label{fig:cortrace}
\end{figure}

\section{Parallax and masses}

From the value of a$\;$sin$\:$i = 1364 $\pm$ 8 Gm (9.118 $\pm$ 0.053
astronomical units) of the spectroscopic orbit and the inclination (i =
$66^{\circ}. 4\pm 0^{\circ}. 5$) of the visual orbit, we obtain the semimajor axis in
AU:
\[
    a = 9.950 \pm 0.069 AU,
\]

\noindent and taking into account the value of the semi-major axis in arcseconds as given
by the visual orbit, we deduce the orbital parallax of the system:
\begin{equation}
    \pi_{orb} = \frac{a''}{a} = 0''.01618 \pm 0''.00023,
	\label{eq:orbp}
\end{equation}

\noindent which is in perfect agreement with the Hipparcos parallax 
$0''.01609 \pm 0''.00065$ \citep{2007AA...474..653V} and which, along with the
new orbital elements, gives a mass sum of 2.357 $\mathcal{M}_{\odot}$.

On the other hand, the spectroscopic orbit provides the mass ratio:
\[
		q = \frac{\mathcal{M}_{1}}{\mathcal{M}_{2}} = 1.079 \pm 0.010,
\]

\noindent while the known expression, with $a$ in AU and $P$ in years:
\[
		\mathcal{M}_{1} + \mathcal{M}_{2} = \frac{a^{3}}{P^{2}} = 2.318 \pm 0.112 \mathcal{M}_{\odot},
\]

\noindent gives the sum of the masses. From that we deduce:
\begin{equation}
\label{eq:masses}
\begin{split}
& \mathcal{M}_{1} = 1.203 \pm 0.059 \mathcal{M}_{\odot}\\
& \mathcal{M}_{2} = 1.114 \pm 0.054 \mathcal{M}_{\odot}.
\end{split}
\end{equation}

The $V$ magnitude of the star, provided by SIMBAD, is 7.37. We can estimate the
difference in magnitude between the components from the speckle
observations made with filters close to that band (2008.7024, 2008.7702,
2010.8920, 2010.9657, and 2013.7369) which would be 0.76.  That would give
us individual magnitudes, m$_{\mathrm{A}}$ = 7.81 and m$_{\mathrm{B}}$ = 8.57.

There are several measurements for the spectral types of the components but
they agree in considering both components to be early--mid-G-type subgiants. 
For instance, \citet{1969AJ.....74..689S} listed them as G3\hspace{2pt}IV--V and,
independently in the same year, \citet{1969PASP...81..643C} measured a
composite spectrum of G2\hspace{2pt}IV, decomposing the system in two stars of the same
type, based on a $\Delta$m of $0^{\mathrm{m}}.0$. SIMBAD adopts the G3/5\hspace{2pt}IV type given in
the {\it Michigan catalogue of two-dimensional spectral types for the HD Stars,
Vol.~5} \citep{1999MSS...C05....0H}.

Using those magnitudes and spectral types, and the Baize--Romani algorithm
\citep{1946AnAp....9...13B} with the parameters given in
\citet{2013MNRAS.428..321D}, we derive a dynamical parallax of 0$''.$01766 for
G2\hspace{2pt}IV + G4\hspace{2pt}IV spectral types, and 0$''.$01630 for G2\hspace{2pt}V + G4\hspace{2pt}V. That is summarized in
Table~\ref{tab:dynpar}. The parallax and the mass sum of the main-sequence model are in better agreement with the values calculated in this work, and also with the Hipparcos parallax and its corresponding mass sum.

\begin{table}
	\centering
	\caption{Parallaxes and mass sums.}
	\label{tab:dynpar}
	\begin{tabular}{lllll}
		\hline
		 & Hipparcos & Orbital & Dynamical & Dynamical \\
		 & & & (Main seq.) & (Subgiant)\\
		\hline
		Parallax & 0$''.$01609 & 0$''.$01618 & 0$''.$01630 & 0$''.$01766 \\
		$\sigma$ & 0$''.$00065 & 0$''.$00023 & 0$''.$00024 & 0$''.$00027 \\
		Mass & 2.357 $\mathcal{M}_{\odot}$ & 2.318 $\mathcal{M}_{\odot}$ & 2.269 $\mathcal{M}_{\odot}$ & 1.784 $\mathcal{M}_{\odot}$ \\
		$\sigma$ & 0.299 $\mathcal{M}_{\odot}$ & 0.112 $\mathcal{M}_{\odot}$ & 0.011 $\mathcal{M}_{\odot}$ & 0.015 $\mathcal{M}_{\odot}$ \\
		\hline
	\end{tabular}
\end{table}

\section{Physical properties of the system}

One of the authors of this study, RFG, has already commented
\citep{2015Obs...135..321G} that these stars are a bit above the main
sequence.  Taking into account the apparent magnitude of the system and the
orbital parallax, the combined absolute magnitude (M$_{\mathrm{v}}$) is 3.40, which
is brighter than expected for a system of main-sequence stars. If we use the
individual magnitudes that we obtained from $\Delta$m, we get M$_{\mathrm{v}}$(A) =
3.84 and M$_{\mathrm{v}}$(B) = 4.60, which are 0$^{\mathrm{m}}$.6 and 0$^{\mathrm{m}}$.4 brighter than the 
calibrated values for the main sequence \citep{1981Ap&SS..80..353S} but
still considerably fainter than for subgiants.

The luminosities of the components calculated from their absolute magnitudes
are 2.36 and 1.17 L$_{\odot}$, respectively.  Interstellar extinction was not
taken into account because it is not expected to be significant at the
distance of the system.  For example, the Geneva--Copenhagen Survey reanalysis
(GCSII, \cite{2011A&A...530A.138C}) gave a value of reddening of 0$^{\mathrm{m}}$.032.

The effective temperatures (T$_{\mathrm{eff}}$) derived from the spectral types are
5794\hspace{2pt}K for the primary and 5598\hspace{2pt}K for the secondary \citep{1981Ap&SS..80..353S}.  We might find several
values of the effective temperature in the literature, obtained by different
methods, but all of them have the problem that they calculate the value for
the whole system, not for each component separately, and this can cause a bias in 
the result. The GCS\hspace{2pt}II gives 6031$\pm$80\hspace{2pt}K but it is included in the 
less-reliable \textit{clbr} sample as the duplicity of the system affects the
photometry. \citet{2012MNRAS.427..343M} used spectral-energy distributions
and atmospheric models to obtain a T$_{\mathrm{eff}}$ = 5378\hspace{2pt}K for a combined
luminosity of 3.89~L$_{\odot}$.  The object is listed
with a T$_{\mathrm{eff}}$ of 5849\hspace{2pt}K in the catalogue of Ages, Metallicities,
Galactic Orbit of F stars \citep{1995BICDS..47...13M}.

\begin{figure}
	\includegraphics[width=\columnwidth]{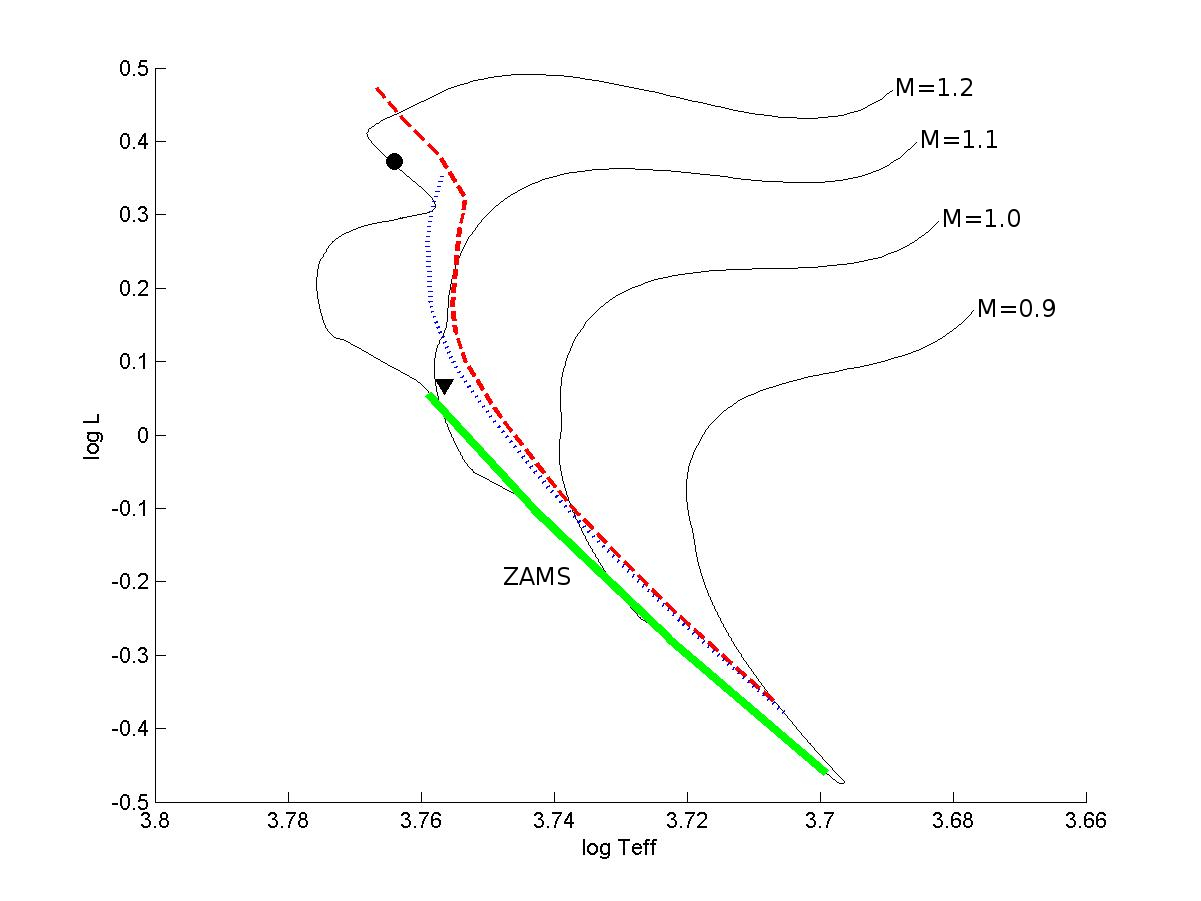}
    \caption{H--R diagram with evolutionary tracks from $\mathcal{M}$ = 0.9 $\mathcal{M}_{\odot}$ to $\mathcal{M}$ = 1.2 $\mathcal{M}_{\odot}$. The dotted line corresponds to the isochrone of 4.47 Gyr and the dashed line to the isochrone of 5.01 Gyr. The circle and the triangle represent the main and secondary components of the system, respectively. The Zero-Age Main Sequence is shown as a thick solid line.}
    \label{fig:evol}
\end{figure}

The metallicity of the system, according to the GCS\hspace{2pt}II, is [\textit{M/H}] = 0.43, so we
can use that value, along with the masses, luminosities, and T$_{\mathrm{eff}}$ to
estimate the age of the system.  We used the PARSEC
\citep{2012MNRAS.427..127B} to generate the isochrones and the evolutionary
tracks with those parameters, and we find the most probable age for the
system to be about 5 Gyr (see Figure~\ref{fig:evol}). 

\section{Conclusions}

We present a new astrometric orbit for the HD 26441 system that is in good agreement with the previous spectroscopic one by RFG in their common elements and has allowed us to calculate accurate masses and the orbital parallax of the system. We have concluded that it is comprised of two G-type stars which are beginning to evolve away from the main sequence, and we have been able to place them in their evolutionary tracks and estimate an age of about 5 Gyr for the system.

\section*{Acknowledgements}

This paper was supported by the Spanish ``Ministerio de Econom\'ia, Industria y Competitividad'' under the Project AYA-2016-80938-P (AEI/FEDER, UE) and by the ``Xunta de Galicia'' under the grant Rede IEMath-Galicia, ED 341DR 2016/022.

\bibliographystyle{mnras}
\bibliography{a2801}

\bsp	
\label{lastpage}
\end{document}